\documentclass{iau}
\usepackage{graphicx}
\usepackage{amsmath}
\usepackage{txfonts}
\begin{document}

\title %% give here short title [Dust/ISM scaling relations]%%
{Intrinsic dust and star-formation scaling relations in nearby galaxies}
\lefttitle{Dust/ISM scaling relations}
\author   %% give here short author list %%
{Bogdan A. Pastrav$^1$}
%%  \thanks{Present address: Fluid Mech Inc., 24 The Street, Lagos, Nigeria.},
% \and Susanne H{\"o}fner$^2$}
\righttitle{Bogdan A. Pastrav}
\affiliation{$^1$Institute of Space Science, \\  Atomistilor 409, 077125 Magurele, Ilfov, ROMANIA
\\ email: {\tt bapastrav@spacescience.ro} \\}%[\affilskip]}

\pubYr{2022} 
\volno{373}  %% insert here IAU Symposium No.
%\pagerange{1--12}
\setcounter{page}{1}
%\jname{Resolving the Rise \& Fall of Star Formation in Galaxies} 
\editors{Tony Wong \& Woong-Tae Kim, eds.}
\jnlDoiYr{2022}
\doival{00.0000/xxxxx}
\aopheadtitle{Proceedings IAU Symposium}

\begin{abstract}
%Accurate dust and star-formation scaling relations are essential in studies of ISM evolution,
%star-formation and galaxy evolution studies, or related to the duty cycle of dust and gas in galaxies.
Following from our recent work, we present results of a
detailed analysis of a representative sample of nearby galaxies. 
The photometric parameters of the morphological components are obtained from bulge-disk decompositions,
using GALFIT software. The previously obtained method and library of numerical corrections for dust,
decomposition and projection effects, are used to correct the measured (observed)
parameters to intrinsic values. Observed and intrinsic galaxy dust and 
star-formation related scaling relations are presented, to emphasize the scale of
the biases introduced by these effects. To understand the extent to which star-formation 
is distributed in the young stellar disks of galaxies, star-formation connected relations which rely 
on measurements of scale-lengths and fluxes / luminosities of H$\alpha$ images, are shown. The 
mean dust opacity, dust-to-stellar mass and dust-to-gas ratios of the sample, together with the main 
characteristics of the intrinsic relations are found to be consistent with values found in the literature. 
\end{abstract}
\maketitle
%\firstsection % if your document starts with a section, remove some space above using this command.
\section{Introduction}
Accurate dust and star-formation scaling relations are essential in studies of interstellar medium (ISM)
evolution, star-formation and galaxy evolution studies, or related to the duty cycle of dust and gas in 
galaxies. These relations can be affected by numerous biases. Among these, it is known that dust introduces
the most significant effects and degeneracies in the observed 
(measured) photometric and structural parameters of galaxies and their main constituents - discs and bulges,
with inclination (projection) and decomposition effects having a non-negligible contribution. These effects 
are stronger at shorter wavelengths and higher disc inclinations (Tuffs et al. 2004,
Pastrav et al. 2013a,b). Obtaining intrinsic relations and show the extent of biases
introduced in their characteristic parameters (zero-point, slope, correlation coefficient), is thus very 
important for the previously mentioned studies.
\section{Method}
A more detailed and complete description of the method is given in \cite{Pas20}. Here we only
briefly summarise the main steps. For this study we considered the images of all the unbarred late-type galaxies 
and lenticulars included in the SINGS (\textit{Spitzer} Infrared Nearby Galaxies Survey; Kennicutt et al. 2003)
and KINGFISH (Key Insights on Nearby Galaxies: a Far-Infrared Survey with \textit{Herschel};
Kennicutt et al. 2011) surveys, in B band and for H$\alpha$ line - in total of 24 galaxies. The B band images were already 
analysed in \cite{Pas20} and \cite{Pas21}. All the images were extracted from the NASA/IPAC Infrared Science
Archive (IRSA) and NASA IPAC Extragalactic Database (NED). 
%NGC0024, NGC0628, NGC1377, NGC1482, NGC1705, NGC2841, NGC2976, NGC3031, NGC3190, NGC3621, NGC3773, NGC3938, 
%NGC4254, NGC4450, NGC4594, NGC4736, NGC4826, NGC5033, NGC5055, NGC5194, NGC5474, NGC5866, NGC7331 and NGC7793.
We then used GALFIT data analysis algorithm (Peng et al. 2010), for the structural analysis of this
sample (bulge-disk decomposition). To fit the disc and bulge surface brightness
profiles, we used the exponential and S\'{e}rsic functions available in GALFIT, and the "sky" function for an initial 
estimation of the image background. A complex star-masking routine was used and the integrated fluxes of the galaxies and
their main constituents were also calculated, together with all the photometric parameters of interest (see Pastrav 2020).
In the next step, central face-on dust opacities ($\tau_{B}^{f}$, $\tau_{H\alpha}^{f}$) for B band and H$\alpha$
line were derived using the empirical correlation found by Grootes et al. (2013), between the face-on dust opacity
and stellar mass surface density (see Eq. (5)). This parameter is essential when determining and applying the
corrections for dust effects. Further on, the dust masses were calculated with Eqs. (2), (3) in \cite{Gro13}
(see also Eqs. A1-A5), which rely on \cite{Pop11} model for the dust geometry, model which takes into
account \cite{Draine03} dust model. % and considers the diffuse dust in the disk distributed axisymetrically in two exponential disks.\\
Next, the chain of numerical corrections for projection, dust \& decomposition effects and the method described in 
\cite{Pas13a,Pas13b} were applied to the measured parameters to derive all the intrinsic parameters involved
in the analysed dust/ISM and star-formation scaling relations. Subsequently, the dust-to-stellar mass ratios and dust-to-gas ratios (observed and corrected) were calculated, using stellar and gas masses found in the literature (e.g. R\'{e}my-Ruyer et al.
2014 \& Grossi et al. 2015). %, Kennicutt et al. 2008 \& Moustakas et al. 2010 - H$\alpha$ fluxes, etc.
\section{Results}
\begin{figure}[h]
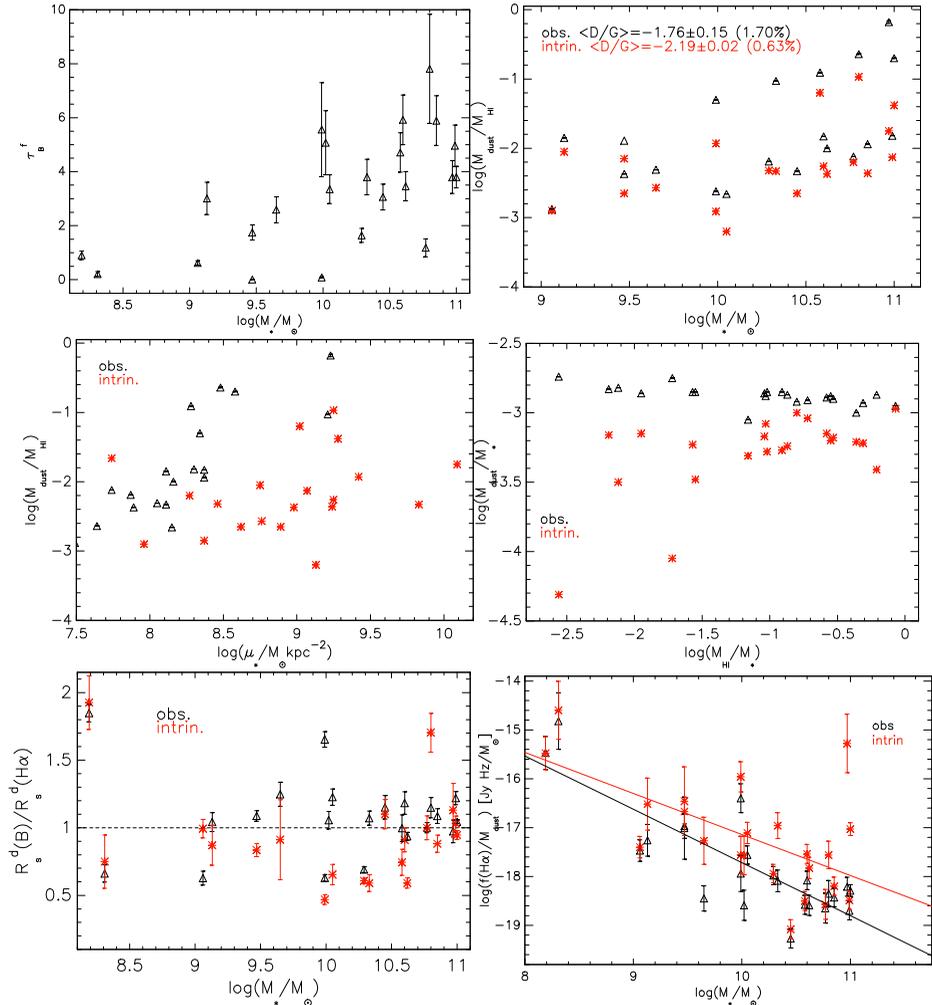

 \begin{center}
  \includegraphics[scale=0.35]{pastrav_fig_1a.epsi}
  \hspace{-0.20cm}
  % \includegraphics[scale=0.35]{dust_opacity_vs_ustar_B.epsi}
   %\vspace{-0.10cm}
   \includegraphics[scale=0.35]{pastrav_fig_1b.epsi}
  % \vspace{-0.10cm}
  % \hspace{-0.20cm}
   \includegraphics[scale=0.35]{pastrav_fig_1c.epsi}
   \hspace{-0.20cm}
   %\vspace{-0.20cm}
   \includegraphics[scale=0.35]{pastrav_fig_1d.epsi}
   %\vspace{+0.50cm}
  %\includegraphics[scale=0.35]{stellar_mass_surf_dens_vs_stellar_mass_B_vSpLen.epsi}
  % \vspace{+1.50cm}
  % \hspace{+0.10cm}
   \includegraphics[scale=0.355]{pastrav_fig_1e.epsi}
   \hspace{-0.10cm}
   \vspace{-0.35cm}
   \includegraphics[scale=0.35]{pastrav_fig_1f.epsi}
  \caption{\label{fig:scal_relations_ISM} Dust/ISM and star-formation related scaling relations. \textit{Upper row} - B band 
  central face-on dust optical depth vs. galaxy stellar mass ($M_{\star}$) \& dust-to-HI ratio ($M_{dust}/M_{HI}$) vs.
  $M_{\star}$; \textit{Middle row} - $M_{dust}/M_{HI}$ vs. stellar mass surface density ($\mu_{\star}$) \& dust-to-stellar
  mass ratio $M_{dust}/M_{\star}$ vs. neutral gas to stellar mass ratio $M_{HI}/M_{\star}$; \textit{Lower row} - B band and 
  H$\alpha$ line emission disc scale-lengths ratio vs. $M_{\star}$ \& the ratio of H$\alpha$ flux to $M_{dust}$ vs. $M_{\star}$.
  Observed quantities are plotted with black triangles and the intrinsic ones with red stars. The error bars show the standard
  deviations, which for certain values are of the size of data points.}
 \end{center}
\end{figure}
We show in Fig.~1 some of the scaling relations derived and analysed in this study. A more complete
discussion will be presented in a forthcoming paper. On the upper row, in the lefthand plot one can
see an increase in $\tau_{B}^{f}$ (similar trend for H$\alpha$ line) with stellar mass $M_{\star}$,
as also recently found by \cite{vanG22} analysing a much larger sample of galaxies from SDSS \& GAMA surveys. 
The derived average $\tau_{B}^{f}$ of the sample, $3.18\pm0.44$, is consistent with studies done on much larger samples 
(e.g Driver et al. 2007 found $3.8\pm0.7$; van der Giessen et al. 2022 found ~4.1), while our dust mass
values are within the error limits of those derived by \cite{Remy14} and \cite{Ani20} using completely different methods.
The increasing trends in the $M_{dust}/M_{HI}$ vs $M_{\star}$ (upper right plot, with $M_{HI}$ being the neutral hydrogen gas mass,
taken from the literature) and $M_{dust}/M_{\star}$ vs $M_{HI}/M_{\star}$ (middle right) is recovered after applying the corrections,
with the average value for dust-to-gas ratios of -2.19 (or 0.63\%) consistent with the one found by \cite{Cort12} of -2.1 for
HI normal galaxies; the rms for the former is low (0.15 - obs. \& 0.03 - intrin.) while the corresponding correlation coefficient
r=0.55 (obs)/0.36 (intrin) is consistent with $r=0.31$ found in \cite{Cort12}. Practically no correlation is observed for the corrected
$M_{HI}/M_{dust}$ vs stellar mass surface density $\mu_{\star}$ (middle left plot), in line with what was found by \cite{Cort12} - 
r=0.07 vs 0.05. A weaker correlation is found for the $M_{dust}/M_{\star}$ vs $M_{HI}/M_{\star}$ relation, with $r<0.5$.
Another scaling relation is shown in the lower right panel, the ratio between the integrated $H\alpha$ flux (or luminosity) and $M_{dust}$ vs.
$M_{\star}$, a consequence of $SFR-M_{\star}$ and $M_{dust}-M_{\star}$ relations, with a derived slope of $-1.09\pm0.16$ (obs.)
/ $-0.95\pm0.20$ (intrin). In the lower left plot, the ratio of the intrinsic disc scale-lengths seen in optical B band and in $H\alpha$ line
as a function of stellar mass is shown, with values generally lower than 1.0 as expected, only after applying the numerical corrections.
A mildly increasing trend can be observed. This shows the extent of optical emission in the disc compared with the extent to which star-formation is distributed
in the young stellar discs of galaxies. \\
Our study underlines the importance of having accurate, unbiased scaling relations in models and studies of ISM evolution and star-formation.
%a) the slope of the B band Mdust-M* relation (Fig. a) is $1.05\pm0.23$ for the observed relation and $0.90\pm0.13$ for the intrinsic one;
%; σ(rms)=0.07 and r(Pearson coefficient)=0.99 for the observed relation; 
%c) the expected decreasing trends with stellar mass or $\mu*$ for the Mdust/M* ratio (Figs. d, e) are only recovered after applying
%the corrections; the average value for the Mdust/M* ratio (-2.88 in log scale, for the observed rel.) is in very good agreement
%with values found of -2.85 by Skibba et al. (2011) or -3.03 by Calura et al. (2017), using other methods;\\
%d) σ=0.34 and r=-0.86 for intrinsic Mdust/M*-$\mu*$ relation, consistent with values of σ=0.45 and r=-0.79 derived by Cortese et al.
%(2012) for galaxies including the KINGFISH spirals; a weaker increasing trend and correlation is found for the $\mu*$-M* relation, 
%with r=0.53 (obs.)/0.57(intrin.) and σ=0.04 dex (obs. and intrin. relations);\\


\begin{thebibliography}{}
\bibitem[Aniano et al. (2020)]{Ani20} {Aniano, G.}, Draine, B. T., Hunt, L. K., Sandstrom, K., Calzetti, D. et al. 2020, ApJ 889, 150
\bibitem[Cortese et al. (2012)]{Cort12} {Cortese, L.}, Ciesla, L., Boselli, A. et al. 2012, \textit{A\&A}, 540, A52
\bibitem[Draine (2003)]{Draine03} {Draine, B.T.} 2003, \textit{ARAA}, 41, 241
%\bibitem[Gadotti et al. (2010)]{Gad10} {Gadotti A. D.}, Baes M., Falony S. 2010, \textit{MNRAS}, 403, 2053
\bibitem[Grootes et al. (2013)]{Gro13} {Grootes, M.}, Tuffs, R.J., Popescu, C.C. et al. 2013, \textit{ApJ}, 766, 59  %\etal
\bibitem[Grossi et al. (2015)]{Grossi15} {Grossi, M.}, Hunt, L. K., Madden, S. C. et al. 2015, \textit{A\&A}, 574, A126
\bibitem[Kennicutt et al. (2003)]{Ken03} {Kennicutt, R. C.}, Armus, L., Bendo, G. et al. 2003, \textit{PASP}, 115, 928
\bibitem[Kennicutt et al. (2011)]{Ken11} {Kennicutt, R. C.}, Calzetti, D., Aniano, G. et al. 2011, \textit{PASP}, 123, 1347
\bibitem[Pastrav et al. (2013a)]{Pas13a} {Pastrav, B. A.}, Popescu, C. C., Tuffs, R. J., Sansom, A. E., 2013a, \textit{A\&A}, 553, A80
\bibitem[Pastrav et al. (2013b)]{Pas13b} {Pastrav, B. A.}, Popescu, C. C., Tuffs, R. J., Sansom, A. E. 2013b, \textit{A\&A}, 557, A137
\bibitem[Pastrav (2020)]{Pas20} {Pastrav, B. A.} 2020, \textit{MNRAS}, 493, 3580
\bibitem[Pastrav (2021)]{Pas21} {Pastrav, B. A.} 2021, \textit{MNRAS}, 506, 452
\bibitem[Peng et al. (2010)]{Peng10} {Peng, C. Y.}, Ho, L. C., Impey, C. D., Rix, H.-W. 2010, \textit{AJ}, 139, 2097
\bibitem[Popescu et al. (2011)]{Pop11} {Popescu, C. C.}, Tuffs, R. J., Dopita, M. A. et al. 2011, \textit{A\&A}, 527, A109
\bibitem[R\'{e}my-Ruyer et al. (2014)]{Remy14} {R\'{e}my-Ruyer, A.}, Madden, S.C., Galliano, F., Galametz, M., Takeuchi, T. T. et al. 2014, \textit{A\&A} 563, A31
\bibitem[R\'{e}my-Ruyer et al. (2015)]{Remy15} {R\'{e}my-Ruyer, A.}, Madden, S. C., Galliano, F. et al. 2015, \textit{A\&A}, 582, A121
%\bibitem[Skibba et al. (2011)]{Ski11} {Skibba, R. A.}, Engelbracht, C. W., Dale, D. . et al. 2011, ApJ, 738, 89
\bibitem[Tuffs et al. (2004)]{Tuf04} {Tuffs, R. J.}, Popescu, C. C., V\"{o}lk, H. J., Kylafis, N. D., Dopita, M. A. 2004, A\&A, 419, 821
\bibitem[van der Giessen et al. (2022)]{vanG22} {van der Giessen, S. A.}, Leslie, S. K., Groves, B., Hodge, J. A., Popescu, C. C. et al. 2022, \textit{A\&A} 662, A26
%\bibitem[Weingartner \& Draine (2001)]{Wei01} {Weingartner, J.C. \& Draine, B.T.} 2001, \textit{ApJ}, 548, 296
\end{thebibliography}
\end{document}